\documentclass[10pt, conference]{IEEEtran}
\usepackage[]{todonotes}
\usepackage[]{booktabs}
\usepackage[caption=false]{subfig}
\usepackage[]{colortbl}

\usepackage{graphicx}
\usepackage{hyperref}

\usepackage{array}

\newcolumntype{R}{%
  >{\tablist}%
  m{3.25in}%
  <{\endtablist}%
}
\newenvironment{tablist}
 {\list{}{%
    \leftmargin=1em
    \itemindent=-1em
    \listparindent=-1em
    \topsep=0pt
    \parsep=0pt
    \partopsep=0pt}\item[\strut]}
 {\endlist\vspace{-\baselineskip}}

\begin{document}

\title{What to Fix? Distinguishing between design and non-design rules in automated tools}
\author{
\IEEEauthorblockN{Neil A. Ernst, Stephany Bellomo, Ipek Ozkaya, Robert L. Nord}
\IEEEauthorblockA{Carnegie Mellon University Software Engineering Institute\\
Pittsburgh, PA\\
Email: \{nernst,sbellomo,ozkaya,rn\}@sei.cmu.edu}}

\maketitle

\begin{abstract}
Technical debt---design shortcuts taken to optimize for delivery speed---is a critical part of long-term software costs. Consequently, automatically detecting technical debt is a high priority for software practitioners. Software quality tool vendors have responded to this need by positioning their tools to detect and manage technical debt. While these tools bundle a number of rules, it is hard for users to understand which rules identify design issues, as opposed to syntactic quality. This is important, since previous studies have revealed the most significant technical debt is related to design issues. 
Other research has focused on comparing these tools on open source projects, but these comparisons have not looked at whether the rules were relevant to design.
We conducted an empirical study using a structured categorization approach, and manually classify 466 software quality rules from three industry tools---CAST, SonarQube, and NDepend. We found that most of these rules were easily labeled as either not design (55\%) or design (19\%). The remainder (26\%) resulted in disagreements among the labelers. Our results are a first step in formalizing a definition of a design rule, in order to support automatic detection.
\end{abstract}

\begin{IEEEkeywords}
Software quality, software design, software cost
\end{IEEEkeywords}

\section{Introduction}
Static analysis tools evaluate software quality using rules that cover many languages, quality characteristics, and paradigms~\cite{Johnson:2013} . For example, the SonarQube tool\footnote{sonarqube.com} can handle C, C++, Java, Javascript, and many others, using language-dependent rule sets (called quality profiles). Software quality rules have accumulated as practitioners gradually recognize code `smells' and poor practices. The first tool to automate rule-checking was the C language tool \texttt{lint} in 1979~\cite{lint}. Today most languages have linters which extend the standard syntax and type checking efforts of a compiler/interpreter to warnings and smell detection. Static analysis tools have traditionally focused on what we might call the \emph{code-level}, rather than design. For instance, Johnson's initial description of \texttt{lint} mentions type rules, portability restrictions, and ``a number of wasteful, or error prone, constructions which nevertheless are, strictly speaking, legal'' \cite{lint}.

Increasingly, however, quality rules are targeting design problems, such as paradigm violations or architecture pattern violations (e.g., the work of Aniche et al. on MVC frameworks \cite{aniche2016validated}). This is because design problems are often more significant than coding errors for long-term software maintenance costs. This view is supported by the results of recent survey and interview \cite{ernst2015measure} and issue tracker analysis \cite{msr/BellomoNOP16} studies we performed, which found that  syntax and coding problems are rarely important sources of these long-term costs, which we call technical debt; instead, sources of debt are created by poor design choices.

The challenge for users of static analysis tools is making sense of the results \cite{Johnson:2013,Bessey:2010}. Static analysis tools often generate many false positives, leading developers to ignore the results \cite{sadowski15}. One potential improvement to this problem is to separate \emph{design rules} from other rules. We examine the software quality rules of three typical tools (CAST, SonarQube, and NDepend) to understand the extent to which their quality rules are design-related. This raises the question of what we mean by \emph{design}, a thorny question in software engineering research \cite[p.14]{Kazman2016}.  We use an extensional definition \cite[ch 1.1]{sep-possible-worlds} of design by creating a design rule classification rubric over several iterations, using rater agreement on classification labels as our metric. Design is clearly more than what the (imperfect) rubric suggests. Most importantly, it is limited to the rules we used as input, and each rater's understanding of design. 
We expand on this notion of design in Section \ref{sec:back}, below.

Our contributions and findings in this work include:
\begin{itemize}
  \item A conceptual model for thinking about design problem detection.
  \item A classification rubric for evaluating design rules.
  \item Existing tools do have rules that can check for design quality. 
  19\% of the rules we examined were design-related.
  \item The rules we analyzed included examples of complex design concepts, such as design pattern conformance and run-time quality improvements. 68\% of the rules that were labelled as design rules were examples of such rules.
\end{itemize}

\section{Background}
\label{sec:back}

\subsection{Separating Design from Non-Design Rules}
\label{sec:drcat}

\begin{figure}[tb]
  \centering
  \includegraphics[width=.35\textwidth]{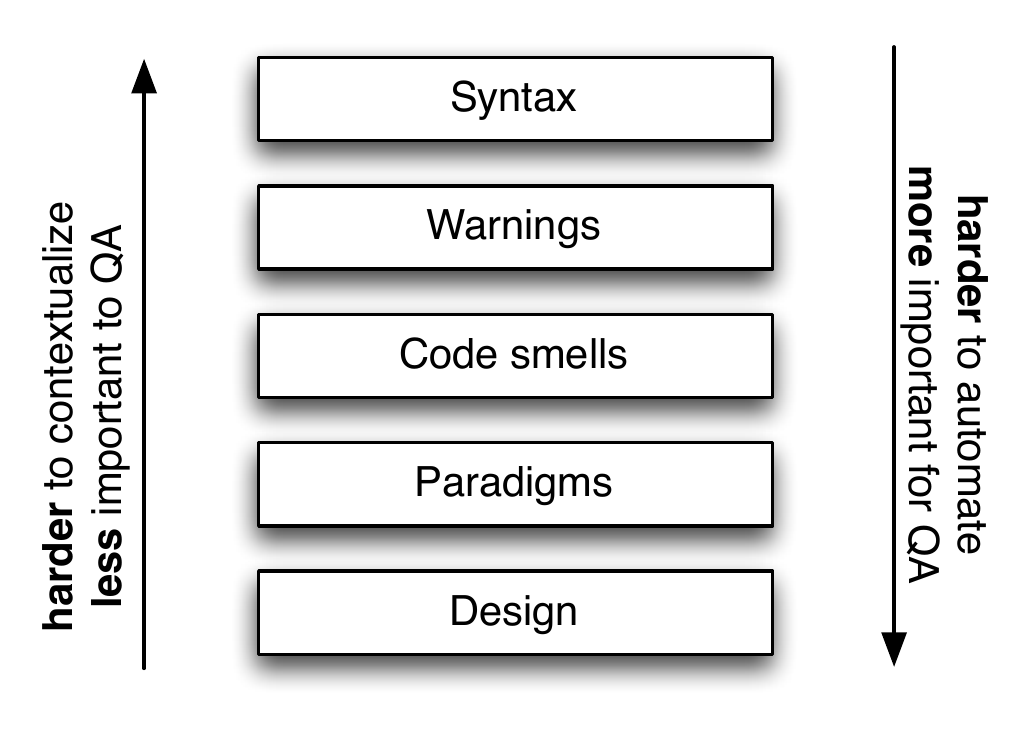}
  \caption{Visual depiction of levels of quality analysis.}
  \label{fig:taxonomy}
\end{figure}

Our goal is to be able to label a given quality rule as either design or non-design. This requires a rubric, a schema that can guide labeling (in the quantitative context, this would be a decision tree). To create this rubric, we began by bootstrapping several concepts of design into  an initial taxonomy, shown in Fig. \ref{fig:taxonomy}. This we based on our experience from conducting numerous software architecture assessments, as well as related literature. This gave a rough classification of design as the upper levels of that taxonomy, i.e., levels ``Paradigm" and ``Design". We defined each level as follows:

\begin{enumerate}
\item
  \textbf{Syntax}---the interpreter or compiler's components (parsers, lexers, intermediate
  forms) assess language/configuration file syntax correctness. (Note that for some specialized cases, levels below this exist: to optimize compiled code with assembler, for example.)
\item
  \textbf{Warning}---compiler warnings (e.g., \emph{--wall}. These warnings tend to be close to syntax
  in their complexity. For example, technically a fall through switch
  statement is syntactically correct in Java, but there is the
  \texttt{-Xlint:fallthrough} tag to catch this and force a compile failure. Often IDEs such as
  Eclipse will flag these automatically with warning icons. Linters
  traditionally operate at this level of analysis.
\item
  \textbf{Code smells}---Code conforms to commonly accepted best
  practices for that language (e.g., for Java, visibility modifiers are
  suitable, in C, no buffer overflows, memory is released appropriately).
  Some cross-language practices apply: documentation, tested, and so on.
\item
  \textbf{Paradigm}---Would someone writing object-oriented,
  functional, embedded, etc. code consider this reasonable? Includes
  principles like SOLID, functional side effects, memory management,
  etc. Distributed code demonstrates awareness of fundamentals of
  distributed computing. Also includes proper use of language idioms
  (e.g.,~proper use of Javascript callbacks, Ruby blocks). 
\item
  \textbf{Design quality}---Given the knowledge available, the code is
  architecturally appropriate for the applicable quality attribute requirement
  (QAR), for example, modular, performant, and secure. The key here is
  understanding the relevant QARs.
\end{enumerate}

Table \ref{tbl:class-example} shows representative rules from all five levels of quality analysis: Java language specification (for level 1: syntax/compile correctness),
Java warnings in Eclipse (level 2), and SonarQube rules detecting code smells, paradigm suitability, and design quality (levels 3/4/5). We use this taxonomy to bootstrap our initial rubric from Section \ref{sec:rubric1}.

\begin{table*}[]
    \caption{Examples of quality rules and their scope}
    \label{tbl:class-example}
    \centering
    \begin{tabular}{p{3.5in}lll}
    \toprule
     Message                                                                                                        & Level           & QAR             & Classification\\
     \midrule
    $<X>$ cannot be resolved to a type                                                                & Syntax/compiler & -               & ND \\
    Return type missing                                                                                      & Syntax/compiler & -               & ND \\
    Empty statement after `if/else'                                                                      & Warning/XLint   & -               & ND \\
    Deprecated                                                                                                      & XLint           & Maintainability & ND \\
    Overrides                                                                                                       & XLint           & Maintainability & ND \\
    Catch clauses should do more than rethrow                                                 & Good code       & Maintainability & ND \\
    Public methods should throw at most one checked exception                      & Good code       & Maintainability & ND\\
    Threads should not be used where ``Runnables" are expected                     & Paradigm        & Concurrency    &  DR \\
    Abstract classes without fields should be converted to interfaces                  & Paradigm        & Modularity      & DR\\
    Component Balance within 1 S.D. of industry average                                   & Design          & Modularity      & DR\\
    Classes should not be coupled to too many other classes                              & Design          & Modularity     &  DR \\
    \bottomrule
    \end{tabular}
\end{table*}

We used a rubric to separate design from non-design rules. To do so, we used inter-rater agreement on the design/not-design labels to assess rubric performance (i.e., how accurately it captured the rater understanding of the two classes). We then reconciled our labels where there was disagreement, and evolved the rubric to account for the lessons from the reconciliation. We then evaluated the rubric a second time, using different raters then the first iteration. To a first approximation, then, \emph{design} is whatever eventual rubric most improves rater agreement (Cohen's kappa~\cite{cohen1968weighted}) for classifying rules, using appropriate safeguards against bias.

\subsection{State of the Practice}
We give a brief overview of the state of the practice with respect to design rule detection, derived from the static analysis tool literature.

While \texttt{lint} and its relatives (jslint, pylint, FindBugs, etc.) focus on code quality that compilers do not catch, increasingly vendors are offering much more than what linters traditionally supported. These \emph{software quality tools} differ in several ways.
\begin{enumerate}
    \item They take a wider perspective on software quality than code faults, incorporating also aspects of design quality;
    \item They are used in managerial roles to check historical trends, as much as for individual developers to check specific faults. Most typically, software quality tools include time as a significant dimension to track trends. Linters tend to flag individual rule violations for follow-up, but provide little support for trends.
    \item They focus on higher-order analysis. Sadowski et al. \cite{sadowski15} call these `more sophisticated static analysis results' and choose not to focus on them, seeing a higher payoff with simpler rules with clear fixes. However, they are focused on developer interaction where time is critical.
    \item Many times they consist of a portfolio/ensemble approach to software metrics and rules, integrating diverse reports on bugs, rule violations, and observations on correlations between different rules.
\end{enumerate}

Thus a software quality tool has a time-series focus, and pays more attention to code-smells, idioms, and design-related issues. These tools fall into the software analytics space \cite{Bird2015}, an emerging research area that seeks to leverage the large volumes of data collected as part of the development process (e.g., lines of code, bugs closed, team structure and activity). No ``tools'' beside IDEs provide compiler level assistance (Eclipse, for example, flags unknown Java types). Very few tools provide integration with compiler flags (see \href{https://docs.oracle.com/javase/7/docs/technotes/tools/windows/javac.html#xlintwarnings}{Java's compiler flags}\footnote{\href{https://docs.oracle.com/javase/7/docs/technotes/tools/windows/javac.html\#xlintwarnings}{https:/\slash docs.oracle.com\slash javase\slash 7\slash docs\slash technotes\slash tools\slash windows\slash javac.html\#xlintwarnings}}).

Software quality monitoring goals can be classified by looking at tool \emph{coverage}. Coverage measures the extent to which quality aspects of
software can be measured. That is, for given design questions,
what answers can be garnered from tools? This might be called coverage
or cognitive support---how well the tools support the analysis goals users have.
The inverse of this question is to see what answers the tools provide
that are simply not useful, or do not map to any known existing
questions.

Tools that incorporate aspects of design analysis ideally provide reliable, automated, repeatable results to address the following goals.

\begin{itemize}
    \item Find poor architectural decisions and shortcuts and identify refactoring opportunities \cite{ernst2015measure,Bouwers:2011tf}.
    \item Understand when payoff is economically justified \cite{Sullivan1999}.
    \item Find increased numbers of \{defects ~ bugs~ churn\} above baseline (hotspots) \cite{xiao2014titan}.
    \item Understand the trends and rate of change in key indicators (such as lines of code, test coverage, or rule violations) \cite{power2013understanding}.
    \item Provide traceability across architectural tiers, frameworks, and languages \cite{aniche2016validated}.
\end{itemize}

\section{Study Scope}
Context is important to the question of how to use automated tools for finding software quality problems. This is because there are contextual factors that make some of what a tool might do superfluous or even irritating for a given project \cite{Johnson:2013}.

\begin{figure}[ht]
  \centering
  \includegraphics[width=.5\textwidth]{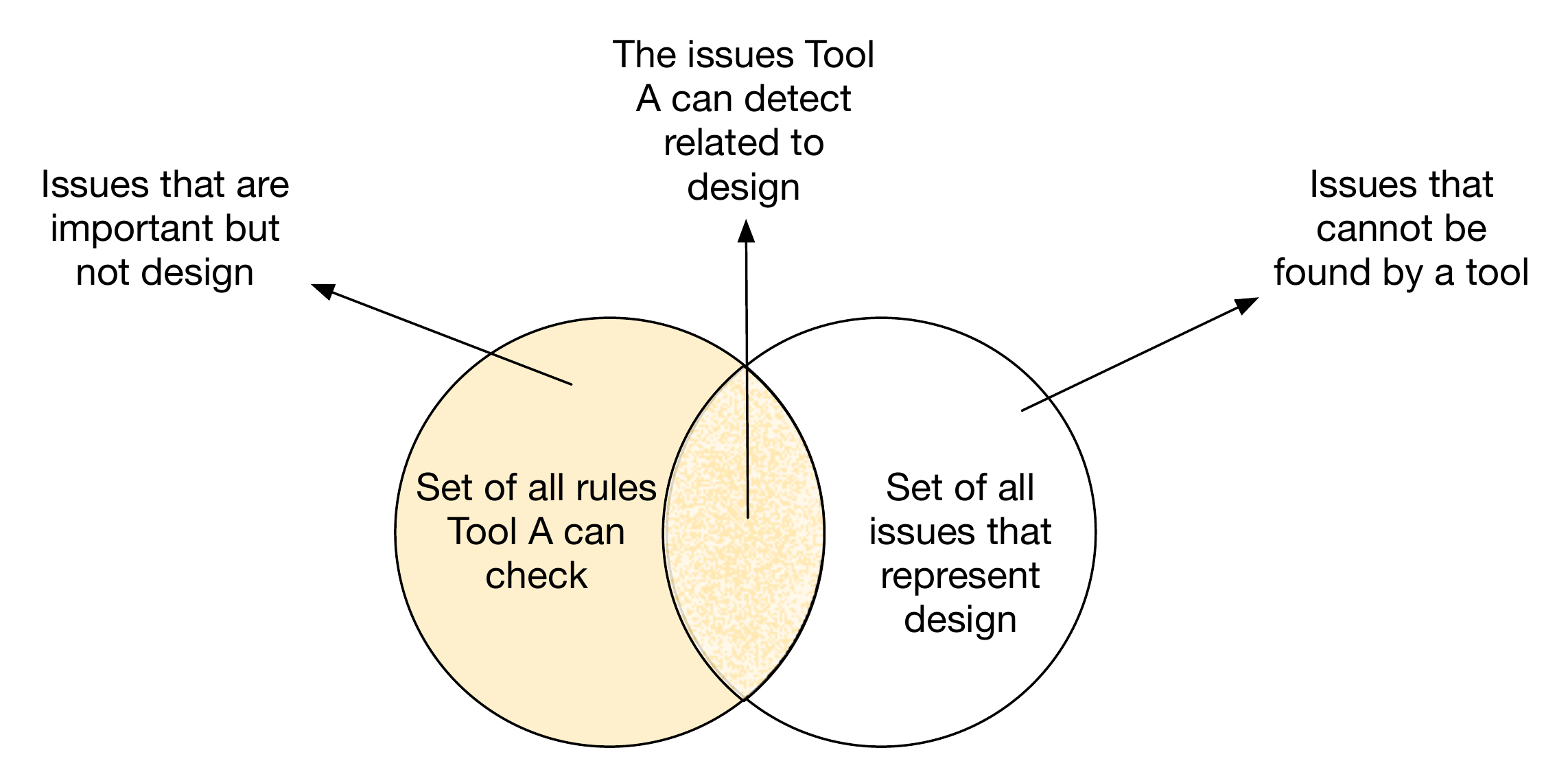}
  \caption{What tools can detect, and what projects want detected. Design issues and rules are both context-dependent on the project properties.}
  \label{fig:conceptual-model}
\end{figure}

Fig. \ref{fig:conceptual-model} shows a conceptual model to organize this space and better articulate our research approach and contributions.

\begin{figure}[tb]
  \centering
  \includegraphics[width=.3\textwidth]{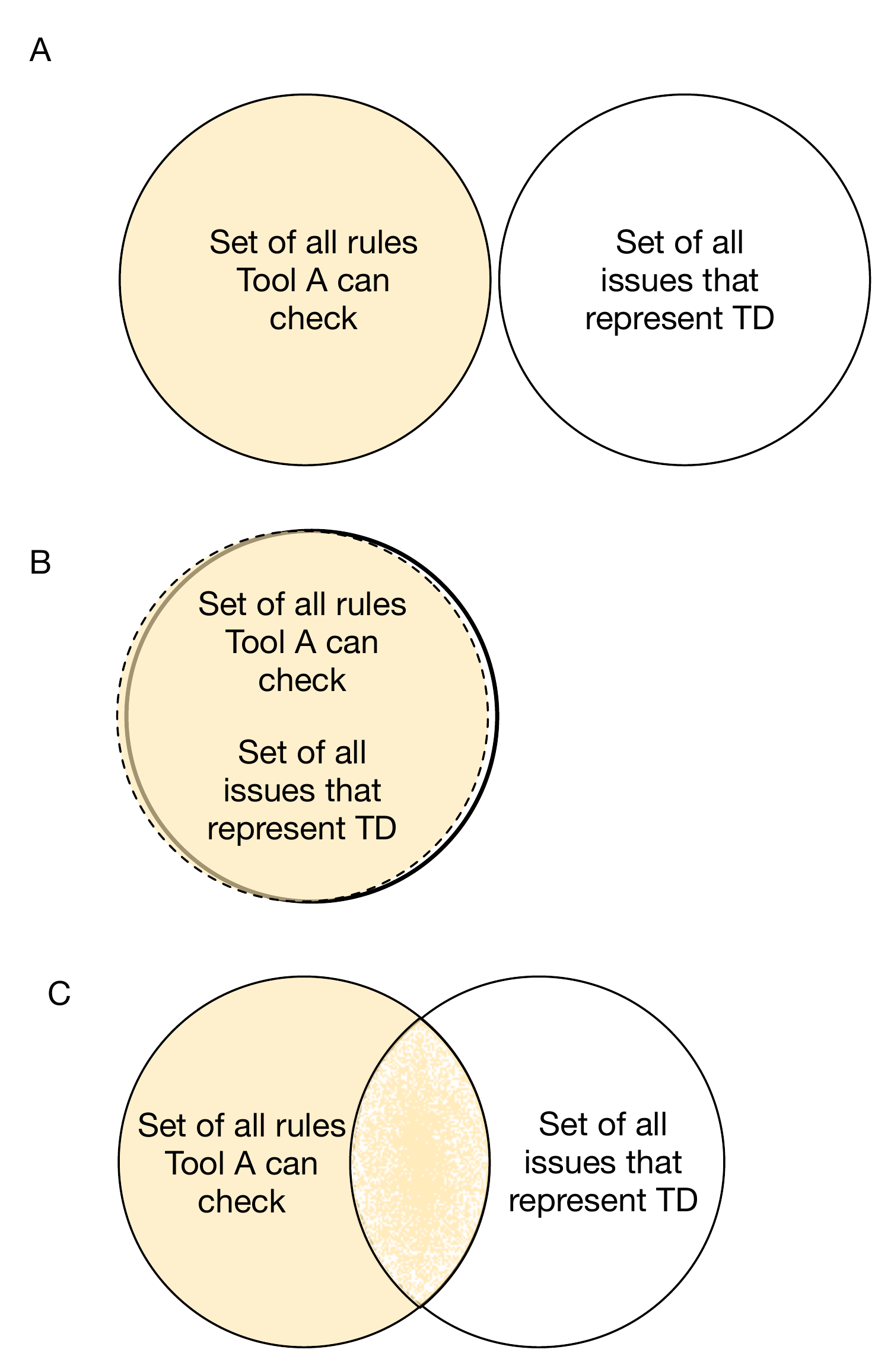}
  \caption{\textbf{A. }The worst case; tools have no capabilities needed. Result: tool goes unused. \textbf{B}. Ideal case. Near-perfect overlap, and the tool is used constantly. \textbf{C}. Reality. The tool has some useful capabilities, but there is a large unmet need.}
  \label{fig:drift}
\end{figure}

Figure \ref{fig:drift} captures our overall research agenda. Modulo the persistent challenge of diffusion of innovations, the goal is to provide insight into moving the two circles closer together. This paper focuses on understanding and improving the automation of design rules. That may be with new tool capabilities (such as usability improvements, new rules, better integration). The other aspect is process change in the organization or project (e.g., moving to continuous integration with automated quality checks).

\section{Methodology}
\label{sec:method}
Our research focuses on how some current tools detect design-related technical debt. This approaches the issue of design problem detection from the left side of Fig. \ref{fig:conceptual-model}; rather than asking what support a given project needs, we look at what support a tool provides to understand design-related rules in the commercial tooling landscape. In particular, we identify two questions:

\begin{itemize}
  \item What is a \emph{design-related} rule?
  \item What kinds of design-related rules do common tools provide?
\end{itemize}
We approached these two questions by evaluating how well three commercial tools manage design rules. We begin with the taxonomy in Section \ref{sec:back}, and then expand that taxonomy into a rubric for classifying rules. We then validate our rubric with experts, and report the rubric performance.

\subsection{Initial Taxonomy}
We began with the taxonomy in Section \ref{sec:back}. The creation of that taxonomy was described previously.

\subsection{Tool and Rule Selection}
We selected the three commercial tools for our study opportunistically. We chose rule sets from software quality management tools that have stated they have capabilities to detect design and technical debt: NDepend\footnote{\url{https://blog.ndepend.com/technical-debt-avoid-ndepend/}}, SonarQube\footnote{\url{https://blog.sonarsource.com/evaluate-your-technical-debt-with-sonar/}}, and CAST\footnote{\url{http://www.castsoftware.com/research-labs/technical-debt}}, and that we have access to (primarily for reasons of licensing and installability). While the tools provide a broader set of capabilities for quality management, we narrowly focused on their quality measures and rules for static analysis of code. In particular, we were able to access all three rule explanations via the tool's documentation.  Then, we chose a subset of available code quality rules. We focused on Java and .Net rules and rules that the tool documentation stated applied generically to all code under analysis. Our analysis of the rules from these tools in no way implies endorsement or critique of the tool itself. 

Each of these tools encapsulate well established code and design quality principles as part of their analysis features.  NDepend provides quality analysis support for .Net applications. NDepend also incorporates some architectural visualization capabilities providing dependency graph and matrix views to interpret system dependencies, element relationships and the like. For NDepend, we analyzed all rules. In NDepend, the rules are grouped into categories, one of which is \emph{Design}. We found that while most of these rules were design-related, several were more about basic principles or metrics, and many rules outside of the design category were still design-related.

SonarQube is an open source platform for code quality analysis that provides support for several languages such as Java, C/C++, Objective-C, Python among others. SonarQube organizes its reports under three categories: bugs, vulnerabilities and code smells and propagates these into a technical debt assessment based on number of days it would take to fix these. For SonarQube, we analyzed all Java rules with priority major or higher. SonarQube rules we analyzed did not explicitly list the categories, such as design or dead code as in NDepend or architecture and application specificity as in CAST.

CAST provides similar features, in addition to dashboards to allow different stakeholders to interpret the results. CAST operates with a service-based model, while it publishes the quality checks its tools embodies through various documentations. The rules we analyzed from CAST were were either JEE or ``all".

\subsection{Rubric Creation and Refinement}
\label{sec:rubric1}
Based on the taxonomy (see Fig. \ref{fig:taxonomy}), empirical data collected in our previous studies \cite{ernst2015measure}, \cite{msr/BellomoNOP16}, and example rules extracted from the three tools, we created our initial design rules classification rubric. The rubric is also motivated by established design and architecture principles, such as assessing the scope of an issue as local, non-local and architectural \cite{Kazman2016}. Such principles are also accepted by tool vendors. For example, CAST's Application Intelligence Platform categorizes its rules into unit level (the rule impacts local part of application code), technology level (the rule impacts several components) and system level (the rule impacts cross boundaries and architectural elements).\footnote{\url{https://goo.gl/6nGEs3}}

We first created a simple definition of a design rule rubric, then iterated on it with examples from interviews and survey responses to `test' how well it handled these cases. The input for the rubric is a single rule from one of the example rule sets. The labeler (person classifying the rule) considers the rule, then applies the decision criteria to the rule. In our classification guidelines we specified that labelers should look at each rule on its own, without considering long-term accumulated impact of multiple violations of the rule. For example, numerous `dead stores' may indicate a bigger problem than a single instance would.

We refined the rubric into a second version, then conducted a final round of classification, rotating assignments so two new labelers approached the dataset. The results reported below apply to this final round.

\subsection{Validation}
We validated the results with arms-length experts in software quality analysis. Each expert applied the rubric to a system they were most recently working on and commented on the extent to which it would correctly distinguish a rule as design-related or not.

\section{Classification Rubric}
\label{sec:rubric}

\begin{figure*}[t]%
    \centering
    \subfloat[]{{\includegraphics[width=.85\textwidth]{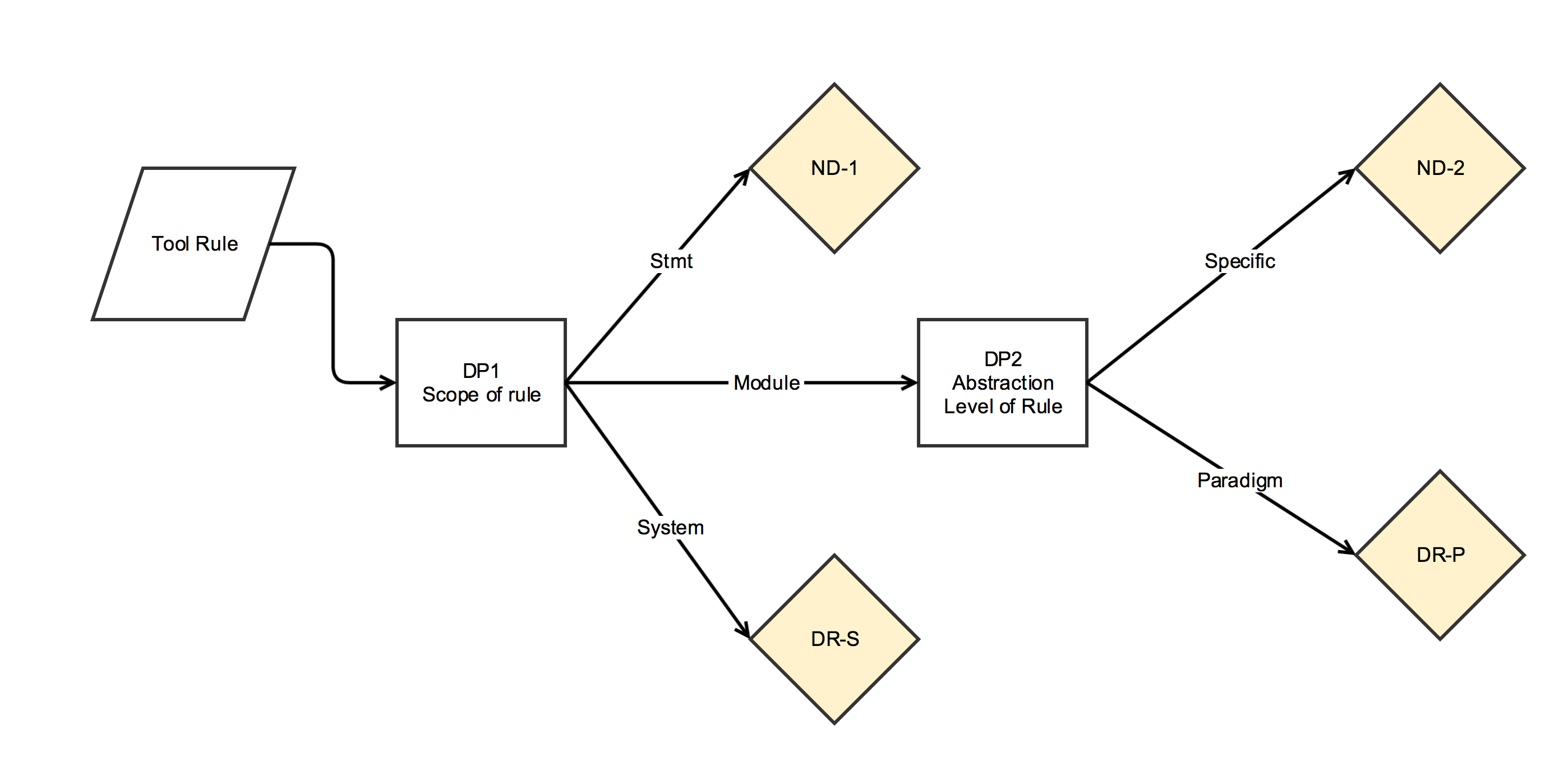}}}%
    \caption{Final version of Design Rule Classification Rubric}%
    \label{fig:design-class}%
\end{figure*}

Following our annotations with the first version of the rubric, we reconciled our labeling results to understand how the rubric failed.  For example, the first version used a quality attribute filter on the rules. We found that the quality attribute filter was not helpful in reducing the label space, since nearly every rule, even syntax rules, had some quality attribute impact. The initial version also had a `deprecated' filter meant to discern design-related rules in the code structure, but was similarly unhelpful; deprecation is rather an orthogonal dimension of the codebase, and a quality rule on its own.

Fig. \ref{fig:design-class} shows our final version of the rubric. The first decision point is rule scope (DP1), with three potential branches that we refer to as statement-level, module-level and system-level below. To explain how it works, we give examples of each branch of the tree below.

\textbf{Statement-level:} At the first filter, statement level, scope is limited to a single code statement and rules are typically syntax related. We categorized as ND-1 (not-design rule) if the rule scope is limited to single code statement (e.g., internal to method (switch, case, if/else, expression). Empty methods, dead stores, also fit here.

We give the following rule as an example here \emph{for loop stop conditions should be invariant}. We labeled this rule ND-1 as the violation is likely to be found within a method, checking for how local variables are set before for loops, a basic coding construct.

\textbf{Module-level:} The next filter is the module-level. This includes groups of statements (that might be bundled into a file or package) or a language construct (method, class) that can be executed independently, reused, tested, and maintained or is a composition of other modules. This category includes module-level syntax specific violations, similar to statement level, as well as rules that check for design paradigms. This leads to a decision point for the abstraction level of the rule (DP2).

We discarded module-level syntax checking rules as non-design, ND-2 in our classification rubric. These rules typically cannot translate easily into another language or paradigm, encapsulate keywords or reserved words or concepts that can't be translated to generalizable concept (e.g., violating Spring naming conventions is a rule with no obvious commonalities in other frameworks). \emph{Don't call GC.Collect() without calling GC.WaitForPendingFinalizers()}
is an example for such rules.

Module-level rules that check for design paradigm, DR-P in our classification rubric, aim to identify rules that encapsulate known design paradigm principles. These include object-oriented, functional, imperative programming, etc.; architectural styles, such as concurrent, model-view-control, pipe-filter, etc.; use of particular design patterns and paradigms, such as exception handling, singletons and factories, etc. \emph{Action Classes should only call Business Classes} is a rule labeled as DR-P since it enforces an aspect of the MVC pattern.

\textbf{System scope:} The system scope filter includes problems detected cross system boundaries (e.g., between languages and/or architectural layers). This also includes rules where system-level metric thresholds are reported (e.g., complexity, dependency propagation). \emph{Avoid having multiple artifacts deleting data on the same SQL table} is an example of such a rule labeled DR-S. This is a design rule because not only multiple system elements are involved, but their architectural responsibilities are critical, in this case enforcing the data model.

Our study artifacts that includes the design rules spreadsheet, categorization labeling results and categorization guidance document (with rubric) are available on a website. \footnote{\url{http://www.sei.cmu.edu/architecture/research/arch_tech_debt/what-to-fix}}

\section{Results}
\subsection{Applying the Rubric to Software Quality Rule Sets}

\begin{figure}[]%
    \centering
\includegraphics[width=.45\textwidth]{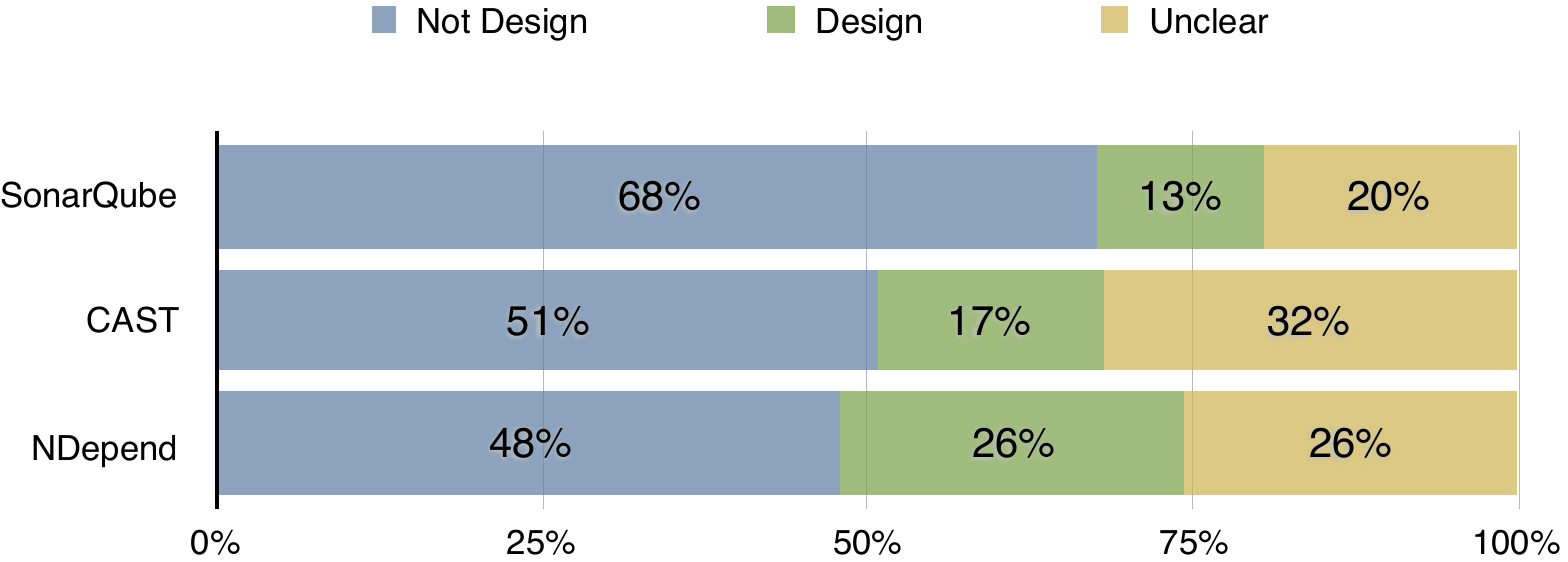}
    \caption{Categories of rater agreement, normalized across tools. Non-design makes up the majority in all three tools.}%
    \label{fig:rubric-relative}%
\end{figure}

Out of the 466 rules we analyzed, 55\% were easily labeled as not design (ND); 19\% were labeled as clearly design, and the remaining had disagreements among the labelers and were hard to classify. Fig. \ref{fig:rubric-relative} shows relative performance of the raters on each tool. In Table \ref{tab:d-nd-examples} we list some examples of design rules, non-design rules and hard-to-classify rules.

 \begin{table}[tb]
  \caption{Examples of Design, Non-Design, and Hard to Classify Rules}
  \label{tab:d-nd-examples}
  \centering
  \begin{tabular}{R}
 \toprule
  \multicolumn{1}{c}{\textit{Design Rules}} \\
   Action Classes should only call Business Classes \\
   Avoid high number of class methods invoked in response to a message \\
   Avoid Classes with a High Lack of Cohesion \\
      \multicolumn{1}{c}{\textit{Non-Design Rules}} \\
   Try-catch blocks should not be nested \\
   Two branches in conditional structure should not have same implementation \\
   All script files should be in a specific directory \\
      \multicolumn{1}{c}{\textit{Hard to Classify}} \\
    Avoid hiding attributes \\
    Avoid defining singleton or factory when using Spring \\
    Avoid declaring an exception and not throwing it \\
   lines of code covered [test coverage] \\
    Each method in an Action Class should have a small complexity \\
    \bottomrule
  \end{tabular}
\end{table}

Table \ref{tab:irr} summarizes the inter-rater agreement (using Cohen's Kappa) after applying this rubric on the data set. For simplicity of reporting, we collapse both design (DR-S and DR-P) and non-design (ND-1 and ND-2) labels together.

\begin{table}[tb]
  \caption{Rater Agreement. DR=Design Rule. ND=Not Design. Gray cells show areas of rater disagreement.}
  \label{tab:irr}
    \centering

\subfloat[NDepend. Cohen $\kappa=0.48. N=148$]{
\begin{tabular}{lc|c}
  & DR  & ND \\
DR & 39 & \cellcolor{gray!25}33  \\ \hline
ND & \cellcolor{gray!25}5 & 71 \\
  \end{tabular}
  } \qquad
\subfloat[SonarQube. Cohen $\kappa=0.44. N=133$]{
  \begin{tabular}{lc|c}
  &  DR  & ND \\
DR & 17 & \cellcolor{gray!25}14\\ \hline
ND & \cellcolor{gray!25}12 & 90  \\
  \end{tabular}} \qquad
\subfloat[CAST. Cohen $\kappa=0.28, N=185$]{
  \begin{tabular}{lc|c}
  &  DR  & ND \\
DR & 32 & \cellcolor{gray!25}24\\ \hline
ND & \cellcolor{gray!25}35 & 94  \\
  \end{tabular}}
  \end{table}

We found it relatively easy to classify many rules on either end of the spectrum as either syntax (e.g., \emph{methods should not return constants} or design (e.g., \emph{action classes should only call business classes}, enforcing the implementation of the MVC pattern.) However, there were many rules that either had characteristics making it hard to classify or where our rubric could be clarified. The low Cohen's $\kappa$ values were due to a high level of disagreement over the hard-to-classify rules.

\subsection{Validation Feedback}
We validated our rubric with three senior architecture analysts from the Software Engineering Institute, not connected with our team.
We had each person comment on the rubric itself, and then label a random sample (n=74) of only the hard to classify rules from the three tools, using the final version of the rubric. We compared each label to a reconciled set of labels that the authors created (where reconciled means we discussed each disagreement to derive a final consensus label). We report Cohen's Kappa and confusion matrices, like that in Table \ref{tab:irr}), in Table \ref{tbl:valid-irr}.

\begin{table}[tb]
  \caption{Validator Agreement. DR=Design Rule. ND=Not Design. Gray cells show areas of rater disagreement.}
  \label{tbl:valid-irr}
    \centering
\subfloat[Validator \#2. Cohen $\kappa=0.25. N=74$]{ 
  \begin{tabular}{lc|c}
  		  &  DR  & ND \\
	   DR & 14 & \cellcolor{gray!25}2\\ \hline
       ND & \cellcolor{gray!25}28 & 30  \\
  \end{tabular}} \qquad
\subfloat[Validator \#3. Cohen $\kappa=0.34, N=74$]{ 
  \begin{tabular}{lc|c}
	&  DR  & ND \\
	DR & 23 & \cellcolor{gray!25}6\\ \hline
	ND & \cellcolor{gray!25}19 & 26  \\
  \end{tabular}} \qquad
\subfloat[Validator \#1. Cohen $\kappa=0.32. N=74$]{ 
\begin{tabular}{lc|c}
  & DR  & ND \\
DR & 21 & \cellcolor{gray!25}5  \\ \hline
ND & \cellcolor{gray!25}21 & 27 \\
  \end{tabular}}
  \end{table}

It is important to note that the validation data set was composed primarily of rules where labelers initially disagreed and then discussed during reconciliation. Our rationale for this criteria was to gather input on effectiveness of classification rubric improvements made after the initial round of classification. While this approach did yield important feedback, the side effect is that focusing on only the challenging labels may have contributed, at least in part, to low agreement numbers.
Following their labeling exercise, we asked the participants for their impressions of the rubric.
\begin{enumerate}
\item \textbf{Scope}. Validators 1 and 3 both expressed difficulty in how to interpret the `scope' filter for its design relevance. Validator 3 explained using an example rule: \emph{Constructors of abstract classes should be declared as protected or private.} Although the method visibility aspect of this rule appeared to have design implications, Validator 3 struggled with the scope aspect explaining, ``I usually defaulted to ND-1 since the change seemed to be limited to a specific statement. However, I wondered should I have said the scope was `system' because of the potential impact across the whole system".
\item \textbf{DR-P/DR-S overlap}. The validators commented on the overlap between the decision branches leading to DR-P and DR-S. For example, a violation that a occurs at one location makes DR-P applicable, however, system boundary implications make DR-S also applicable. Validator 1 said the decision was further complicated by trouble deciphering what it meant for problems to cross system boundaries, ``Is the problem located at multiple points in the system?  Does it affect multiple points...?" (we noted the latter comment is similar to the overarching scope comment given by Validator 3 above).
\item \textbf{Metric threshold rules}. Metric threshold is about rules such as \emph{Module complexity over x limit}. Validator 1 said that scope frequently led him down the ND-2 module path because heuristics are frequently applied at the method or class level in the rules. However, metric thresholds show up as an example in DR-S, too.
\end{enumerate}

\section{Discussion}
Research to date on the fitness of quality analysis tools for design analysis has taken the approach of running multiple tools on the same data sets and comparing the results with each other \cite{falessi2015validating,mtd/GriffithRICDW14}, or with some measure of design ground truth \cite{archrecovery/LutellierCGTRMK15}. Such approaches have limitations due to potential feature limitations of tools. Consequently, they assess the tools and report what cannot be done, but do not improve our understanding of design analysis based on core concepts. Our goal in analyzing the rule sets is to assess the characteristics of automatable rules that check for design problems.

\subsection{The Nature of Design Rules}
Design rules have one or more of the following properties:
\begin{itemize}
  \item Design rules check for propagation of issues across system elements. All of the rules where there was agreement on as design rules had this property.
  \item Design rules encapsulate semantics of known design constructs, such as architectural patterns. While some rules were initially tossed out as syntactic, recognizing that their goal was to enforce design constructs or patterns helped clarify their assessment.
  \item Design rules break down semantics of quality attribute concerns. This is an area where most significant progress can be made. Certain quality attribute concerns can still be automatable and enforce design; exception handling, test coverage were such examples in the rules we analyzed. We see the biggest gap in existing rules in taking explicit advantage of a mapping between automatable analysis and quality attribute concerns and design tactics.
  \item Design rules assess system impact over time, which is tricky as this is where there is no magic number or metric and is observed over time in the context of the system. We did not include reporting rules in this set such as source lines of code, number of functions changed and the like.
\end{itemize}

Exploring the research question: \emph{What kinds of design-related rules do common tools provide?} gave input into the intersection of what the rule sets of common tools can detect, and what design analysis is needed to detect technical debt. We found several problems with the software quality rules the tools provided:
\begin{itemize}
    \item Rules evolved from metrics that are possible to calculate and automate, but those metrics are not necessarily the most useful for the information needs of tool users.
    \item Rules that the user chooses not to take action on are due to false positives, lack of context integration, and overload of data without actionable guidance on what to fix remain a large problem.
    \item Many rules focus on analysis at the `code smells' level and below (taxonomy levels 1-3). Problems at the paradigm suitability and design quality levels (4,5) remain poorly supported. However, this was the source of the greatest amount of technical debt according to our earlier studies \cite{ernst2015measure}, \cite{msr/BellomoNOP16}.
\end{itemize}

\subsection{Lessons Learned in Assessing Rules}
\begin{itemize}
  \item \emph{Are syntactic rules checking for design conformance?} Our rubric led us to classify rules checking purely syntactic implementation as not design related. However, the goal of some syntactic rules is to enforce design conformance. Examples of such rules are ``avoid declaring an exception and not throwing it" or ``classes should not be empty" indicating dead code in some cases. 
  \item \emph{Are metric threshold rules indicative of design problems, and thus design rules?} While a number of software metrics violating a certain threshold are available as rules (e.g., cyclomatic complexity should not exceed 7, depth of inheritance should not exceed 5), there is also evidence that such heuristics have a wide range of false positives and disregard context. These thresholds make sense only when combined and correlated with other system observations. These rules are helpful in creating other complex rules, yet are not useful for design assessment solely by themselves.
  \item \emph{Are reporting rules design rules?} Reporting rules measure source lines of code, class size, method size, line length, number of parameters used, and the like. We concluded that they are not design rules, but need to be treated as contextual parameters that allow improved analysis of systems with relevant subsets of the rules.
\end{itemize}

\subsection{Threats to Validity}
We identified the following threats to validity.

\textbf{Manual inspection:} Our research approach relies on manually classifying existing software quality rules as implemented by commercial tools for their ability to detect design issues. In order to minimize biased classification, we had multiple researchers label the rules, and revised the classification guidance based on the biases we identified. In addition, we also had experts external to the research team classify a random sampling of the issues. The team and external experts classifying the rules are experts in technical debt research and software design, but they are not all experts in the tools or languages studied.

\textbf{External validity:} At this stage of our study we do not claim generalizable results, i.e., we do not expect this rubric, and its associated definition of design, to necessarily work on other tools. We continue to address these open research questions by applying the outcome rule set to several customer and open source projects, which is ongoing work that is not in the scope of this paper.

\textbf{Internal validity:} We checked our inter-rater reliability by having two of us assess rules, and rotating raters after round one. We resolved disagreements and reflected the outcomes of that in the rubric. We only focus on a subset of rules for Java and C\# and those that these particular tools have defined. However, the rules focus on basic language constructs that are transferable and we included rules from several tools to minimize this threat.

\textbf{Construct Validity:} The initial taxonomy relies on our initial view of software design to characterize software quality rules. Other definitions of software design (as opposed to code quality) might result in a different taxonomy. Our rubric relies on our understanding of design, as well as the literature and tool's internal documentation.

\section{Related Work}
\label{sec:rel}
\subsection{Finding Design Violations}
A few approaches, embody architectural design constraints as first class constructs. Koziolek surveyed architecture metrics and categorized a number of quantitative methods \cite{Koziolek2011}. Some studies have looked at validating a collection of measures in a quality model (mostly for maintainability \cite{Ploesch2015}). Others have looked in depth at one or more methods to assess the efficacy of the approach (for example \cite{Bouwers:2011tf}). The Consortium for IT Quality (CISQ) Automated Quality Characteristic Measures are used to identify violations of good coding and architectural practice in the source code of software \cite{CISQ2012}. CAST has rules to implement the measures, for example, it allows one to define constraints on relations between modules, and highlights violations \cite{Curtis2012}. Lattix is specifically targeted to architectural analysis, using static dependency clusters as the unit of analysis \cite{Sangal2005}. Reflexion models were the initial approach to defining this \cite{murphy:1995aa}; since extended into architecture patterns \cite{AnvaariZ14}. The other approach is architecture reconstruction by detecting architectural tactics and design patterns \cite{Mirakhorli2012} or design rule spaces \cite{Xiao2014}.

A number of studies have looked for relationships between software metrics as reported by software quality tools as those we used for our study and technical debt. This work has applied existing code smells, coupling and cohesion, and dependency analysis to identifying areas of technical debt \cite{Fontana:2012}, \cite{falessi2015validating}. The outcomes of these studies correlate with our findings where some subsets of the rules do represent more significant issues in the code base. There is also work that investigates the analysis of technical debt problem as a design flaw issues, such as \cite{Marinescu12/TDDesignFlaw}. Similarly, Kazman et al. relates architectural modularity violations to number of bugs to detect technical debt \cite{Kazman/SoftServTD}. The goal of our study is to create a framework by which we can extend and limit the scope of existing rules to repetably detect design quality. The design issues highlighted in these studies overlap with those rules that our study identified as design rules and represent the modifiability/maintainability space.

Software quality tool vendors are grappling with the same challenges. Recently SonarQube simplified its quality model to create a better encapsulation of code issues, maintainability issues and run-time aspects that are most critical such as vulnerabilities\footnote{https://blog.sonarsource.com/bugs-and-vulnerabilities-are-1st-class-citizens-in-sonarqube-quality-model-along-with-code-smells/}. CISQ recently put out a developer survey to understand how developers perceive the time to fix when they encounter certain violations that such rules tag to better assess technical debt\footnote{http://it-cisq.org/technical-debt-remediation-survey/}.


\subsection{Use of Tools}
Johnson et al. \cite{Johnson:2013} write about interviews they did at Google with twenty developers asking how they used Lint, Findbugs etc. The main findings were it took them out of their workflow and there were too many false positives. A similar perspective was voiced in our survey in terms of using such tools for technical debt analysis \cite{ernst2015measure}.

Emanuelsson and Nilsson \cite{Emanuelsson20085} describe how three static analysis tools are used at Ericsson to find bugs: Coverity, Polyspace, and Klocwork. After describing the different features of each tool, they turn to experiences at Ericsson with the tools. Among their conclusions were that the tools generally performed quite well, with manageable false-positive rates, if the tool was introduced early on. Here ``false-positive'' is used in the \emph{irrelevant} sense to capture the developer's perspective: any report from the tool a user chooses not to take action to resolve the report, rather than the strict sense, namely, a result that should never have occurred. Sadowski et al. \cite{sadowski15} call this `effective false positives'. Falessi and Vögele discuss this distinction as well \cite{falessi2015validating}. The Ericsson result also found that tools were complementary to one another, but desired better tuning and configurability options, e.g., ``a mechanism to fine-tune the effort spent on deriving value ranges of variables and the effort spent on aliasing analysis".

In Smith et al. \cite{Smith2015} and  Witschey et al. \cite{WitscheyZWMMZ15} the authors investigate the usefulness and patterns of work for security analysis tools. Smith et al. examine how developers use security analysis tools. This builds on earlier work that examines information needs for software developers, such as Ko's work \cite{Ko2005} that looked at questions asked in program understanding tasks. Smith et al. took ten developers and assigned them security analysis tool to walk through bug reports, categorizing the types of questions they asked while doing so. Card sorting produced over ten classes of common questions new tools should seek to address. Witschey et al. did a similar exercise looking at what intrinsic properties  of those tools were important for adoption. The differences between the two papers---focused on information needs developers have, and also support tools provide---mirrors the framework we adopt for technical debt tools.

For security tools, Witschey et al. \cite{WitscheyZWMMZ15} created a regression model to predict whether a tool would be adopted. The factors in that model include that one can see what others do with the tool (Observability), frequent organizational training and mandates (Education and Policies), and a perception that security tools are valuable (Advantages). It seems reasonable to think that most of these factors would generalize to other quality attributes and software quality in general.

Smith et al. \cite{Smith2015} sought to understand the information developers need when using a static security analysis tool. The outcome was a table of important information needs, including ``Preventing and Understanding Potential Attacks", and understanding ``Control and Call Information", among several others. The results are mostly specific to security, but a class of needs about developer self-awareness and education are likely useful for technical debt tools as well.

There are also several tools to assist with refactoring. What is relevant to the design analysis and technical debt tool landscape is that despite an established and industry-championed set of practices (``Refactoring", by Martin Fowler \cite{Fowler1999}, is widely considered a standard book in the practice of software engineering), excellent IDE integration, many refactoring practices are poorly adopted. This puzzling state of affairs is investigated by Murphy-Hill and Black \cite{4602672} and in more detail by Murphy-Hill, Parnin, and Black \cite{emerson12}. They note that close to 90 percent of refactorings are performed manually, despite good tool support. Rename was by far the most common. 

In analyzing the tools themselves, Murphy-Hill et al. \cite{emerson12} notes that ``Toolsmiths need to explore alternative interfaces and identify common refactoring workflows'' to better integrate the refactoring with what developers need at that moment. They list three key factors in whether a developer will use the tool: \emph{Awareness} of the tool, \emph{Opportunity} to apply the tool at the right moment, and \emph{Trust} that the tool will do what is needed. The authors also note two reasons developers gave for not using a tool: 1) unfamiliar touch points in the code, those places the refactoring would change; 2) disruption in normal routines by jumping into new UI elements and out of the problem at hand. Both the three factors and two obstacles seem relevant to technical debt tools as well. 

There is a renewed interest in software analytics, and many corporations are adopting them as part of their practice. Sadowski et al. \cite{sadowski15} describe how Google integrate static analysis and technical debt identification into Google's development practices and environments, using a tool called Tricorder (now released as an external open source project called Shipshape). They begin by noting the problems with existing tools: Google's giant codebase; engineers testing their own code and mandatory code reviews for each patch; high false-positive rates with tools in the past, such as Findbugs; and a wide variety of programming languages and contexts. The Tricorder tool addresses this last point by supporting domain-specific analysers using a contract approach, that mandates these plugins not have high false positive rates (that is, report results that developers label annoying or uninformative). 

\section{Conclusion}
\label{conclusion}
Software quality tools mix design rules and code quality rules. Separating these is important since design problems often result in longer-term costs. This study labeled the software quality rules of three tools as \emph{design} or \emph{not-design}, using an iteratively defined classification rubric validated with experts. 

Our study suggests that progress in automated design analysis can be achieved by addressing the following:
\begin{enumerate}
    \item\textbf{Defining design rule scope}. Our classification results revealed design rules go beyond statement level quality checks. Tools that reported scope of impact (and not just time to fix, as many do currently), would aid in classification.
    \item\textbf{Validating properties of design rules}. Based on the design rules we identified, we extracted initial properties of design rules. For example, we found that the quality attribute filter was not helpful since nearly every rule, even syntax rules, had some quality impact. Existing know-how on design tactics and expert observations can help validate and improve the properties of automatable design rules. 
    \item \textbf{Improving context sensitivity}. The hard to label rules in our data are an important outcome of our study. Often, the reason they were hard to correctly label was due to context-specificity. As reported by Microsoft researchers in \cite{zhang2013software}, it is only when your analytics efforts work closely with the information needs of the stakeholders that there is real impact. 
\end{enumerate}

\section{Acknowledgments}
This material is based upon work funded and supported by the Department of Defense under Contract No. FA8721-05-C-0003 with Carnegie Mellon University for the operation of the Software Engineering Institute, a federally funded research and development center.

References herein to any specific commercial product, process, or service by trade name, trade mark, manufacturer, or otherwise, does not necessarily constitute or imply its endorsement, recommendation, or favoring by Carnegie Mellon University or its Software Engineering Institute.

[Distribution Statement A] This material has been approved for public release and unlimited distribution. Please see Copyright notice for non-US Government use and distribution.

DM-0004376

\bibliography{tools.bib}
	\bibliographystyle{IEEEtran}

\end{document}